\documentclass[review]{elsarticle}

\usepackage{hyperref}

\usepackage{epsfig,graphicx,float,color}

\newcommand\rr{\color{black}}

\begin{document}

\begin{frontmatter}

\title{Measuring photometric redshifts using galaxy images and Deep Neural Networks}

\author{Ben  Hoyle}
\address{Universitaets-Sternwarte, Fakultaet fuer Physik, Ludwig-Maximilians Universitaet Muenchen, Scheinerstr. 1, D-81679 Muenchen, Germany}
\address{Excellence Cluster Universe, Boltzmannstr. 2, D-85748 Garching, Germany}

\cortext[mycorrespondingauthor]{Corresponding author}
\ead{hoyleb@usm.uni-muenchen.de,benhoyle1212@gmail.com}

\begin{abstract}
We propose a new method to estimate the photometric redshift of galaxies by using the full galaxy image in each measured band. This method draws from the latest techniques and advances in machine learning, in particular Deep Neural Networks. We pass the entire multi-band galaxy image into the machine learning architecture to obtain a redshift estimate that is competitive, in terms of the measured point prediction metrics, with the best existing standard machine learning techniques. The standard techniques estimate redshifts using post-processed features, such as magnitudes and colours, which are extracted from the galaxy images and are deemed to be salient by the user. This  new method removes the user from the photometric redshift estimation pipeline.  However we do note that Deep Neural Networks require many orders of magnitude more computing resources than standard machine learning architectures, {\rr and as such are only tractable for making predictions on datasets of size $\leq$50k before implementing parallelisation techniques}. 
\end{abstract}

\end{frontmatter}

\section{Introduction}
To maximise the cosmological information available from current and upcoming large scale galaxy surveys, one requires robust distance estimates to many galaxies. The distances to galaxies are inferred by the distance-redshift relation which relates how the galaxy light is stretched  due to the expansion of the Universe as it travels from the galaxy to our detectors. This stretching leads to {\rr an energy} loss of the photon and a shift towards redder wavelengths, which is  known as the redshift.  The further away the galaxy is from us, the longer the light has been passing through the expanding Universe, and the more it becomes redshifted.

Obtaining very accurate spectroscopic redshifts, which measures the redshifted spectral absorption and emission lines,
requires very long exposure times on dedicated spectrographs and is typically only performed
for a small sub-sample of all galaxies.  Conversely, the measurement of multi-band photometric properties of galaxies
is much cheaper. The compromise is then to attempt to extract less accurate redshift information from photometrically
measured properties, but applied to a much larger galaxy sample.

Photometric redshift estimates are obtained from either template fitting techniques, machine learning techniques, or some hybrid of the two for example using data augmentation \cite{2015arXiv150106759H}.  The template methods are parametric techniques and are constructed from templates of the Spectral Energy Distribution of the galaxies. Some templates encode our knowledge of stellar population models which result in predictions for the evolution of galaxy magnitudes and colours. The parametric encoding of the complex stellar physics coupled with the uncertainty of the parameters of the stellar population models, combine to produce redshift estimates which are little better than many non-parametric {\rr techniques. See} e.g., \cite{2010A&A...523A..31H,2013ApJ...775...93D} for an overview of different techniques. Unlike non-parametric and machine learning techniques, the aforementioned template methods do not rely on training samples of galaxies, which must be assumed to be representative of the final sample of galaxies for which redshift estimates are required. Other template methods are generated either completely from, or in combination with, empirical data, however these templates both require tuning, and also rely upon representative training samples.

When an unbiased training sample is available, machine learning methods offer an alternative to template methods to estimate galaxy redshifts. The `machine architecture' determines how to best manipulate the photometric galaxy input properties (or `features') to produce a machine learning redshift. The machine attempts to learn the most effective manipulations to minimise the difference between the spectroscopic redshift and the machine learning redshift of the training sample.

The field of machine learning for photometric redshift analysis has been developing since \cite{2003LNCS.2859..226T} used artificial Neural Networks (aNNs). A plethora of machine learning architectures, including tree based methods, have been 
applied to the problem of point prediction redshift estimation \cite{2014arXiv1406.4407S} or to estimate the full redshift probability distribution function  \cite{2010ApJ...715..823G,tpz,2013arXiv1312.1287B,2015arXiv150308215R}. Machine learning architectures have also had success in other fields of astronomy such as galaxy morphology identification, and star \& quasar separation \cite{1997daa..conf...43L,2009arXiv0910.3770Y}.

The use of Deep Neural Networks (hereafter DNN) as the machine learning architecture has only recently been applied to problems in astrophysics. For example \cite{2015arXiv150307077D} taught a DNN to replicate the detailed morphological classifications obtained by the citizen scientists answering questions within the Galaxy Zoo 2 project \cite{2013MNRAS.435.2835W} and obtained an accuracy of up to $99\%$ on some classification questions, and \cite{2014arXiv1412.8341H} examined the problem of spectral classification from Sloan Digital Sky Survey \cite{2014ApJS..211...17A} (hereafter SDSS) spectra.

Within the standard machine learning approach the choice of which photometric input features to train the machine architecture, from the full list of possible photometric features, is still  left to the discretion of the user.  The current author recently performed an analysis of `feature importance' for photometric redshifts, which uses machine learning techniques to determine which of the many possible photometric features produce the most predictive power \cite{2015MNRAS.449.1275H}. The technique described in this paper is the most extreme example of feature importance possible. We no longer need to impose our prior beliefs upon which derived photometric features produce the best redshift predictive power, or even measure the photometric properties. By passing the entire galaxy image into the Deep Neural Network machine learning framework we completely remove the user from the photometric redshift estimation process. 
 
Furthermore in order to use either the template or standard machine learning techniques to estimate redshifts, the magnitudes, colours, and other properties  of the galaxies must be measured. The analysis presented in this paper, which uses the full image of the galaxy partially removes this requirement. However we do still currently need the galaxy to have been detected so that we can generate a postage stamp image.
 
The outline of the paper is as follows. In \S\ref{data} we describe the galaxy images and the pre-processing steps to prepare the images for the Deep Neural Networks. We then introduce both of the machine learning architectures in \S\ref{MLA}, and present the analysis and results in \S\ref{res}. We conclude and discuss in \S\ref{concl}.

\section{Galaxy data and images}
\label{data}
The galaxy data in this study are drawn from  the SDSS Data Release 10 \cite{2014ApJS..211...17A}. The SDSS I-III uses a 2.4 meter telescope at Apache Point Observatory in New Mexico and has CCD wide field photometry in 5 bands \cite{Gunn:2006tw,Smith:2002pca}, and an expansive spectroscopic follow up program \cite{2011AJ....142...72E} covering $\pi$ steradians of the northern sky. The SDSS collaboration has obtained  2 million galaxy spectra using dual fibre-fed spectrographs. An automated photometric pipeline performs object classification to a magnitude of $r\approx$22 and measures photometric properties of more than 100 million galaxies. The complete data sample, and many derived catalogs such as the photometric properties, and 5 band FITS images are publicly available through the SDSS website\footnote{sdss.org}.

We obtain 64,647 sets of images from the SDSS servers for a random selection of galaxies which are chosen to pass the following photometric selection criteria; the angular extent must be less than 30 arc seconds as measured by the `Exponential' and {\rr `de' Vaucouleurs'} light profiles in the $r$ band; and that each $g,r,i,z$ have magnitudes greater than 0. We further select galaxies which pass the following spectroscopic selection criteria; the error on the spectroscopic redshift to be less than 0.1 and the spectroscopic redshift must be below 2. {\rr We check that none of the selected galaxies have images with missing or masked pixel values. In detail we run the MySQL query as shown in the appendix in the CasJobs server.}

We choose to obtain the galaxy image FITS files in the following four photometric bands; $g,r,i,z$. This enables a closer resemblance to the bands available in other photometric surveys, for example the Dark Energy Survey\cite{2005astro.ph.10346T}.  Each pixel in the FITS file has a resolution of 0.396 arc seconds and contains the measured flux which has been corrected for a range of observational and instrument effects such as flat fielding and sky subtraction, in order to be suitable for astronomical analysis. {\rr All pixel fluxes are converted to pixel magnitudes following \cite{1999AJ....118.1406L}. We apply a further extinction correction to account for galactic dust  using the maps of \cite{1998ApJ...500..525S} which is available from the photoObjAll table in the CasJobs server. The extinction corrections are subtracted from the value of magnitude in each pixel} in the corresponding FITS files. We choose to use FITS images of size 72x72 pixels, corresponding to 28.5 arc seconds on a side. We have explored the use of other image dimensions (32x32) but do not find improvement in the obtained results. {\rr The chosen image size is motivated by, and closely follows earlier work using SDSS images \citep[][]{2015arXiv150307077D}, and ensures that the training times are tractable.}

\begin{figure}[ht]
\includegraphics[width=400px]{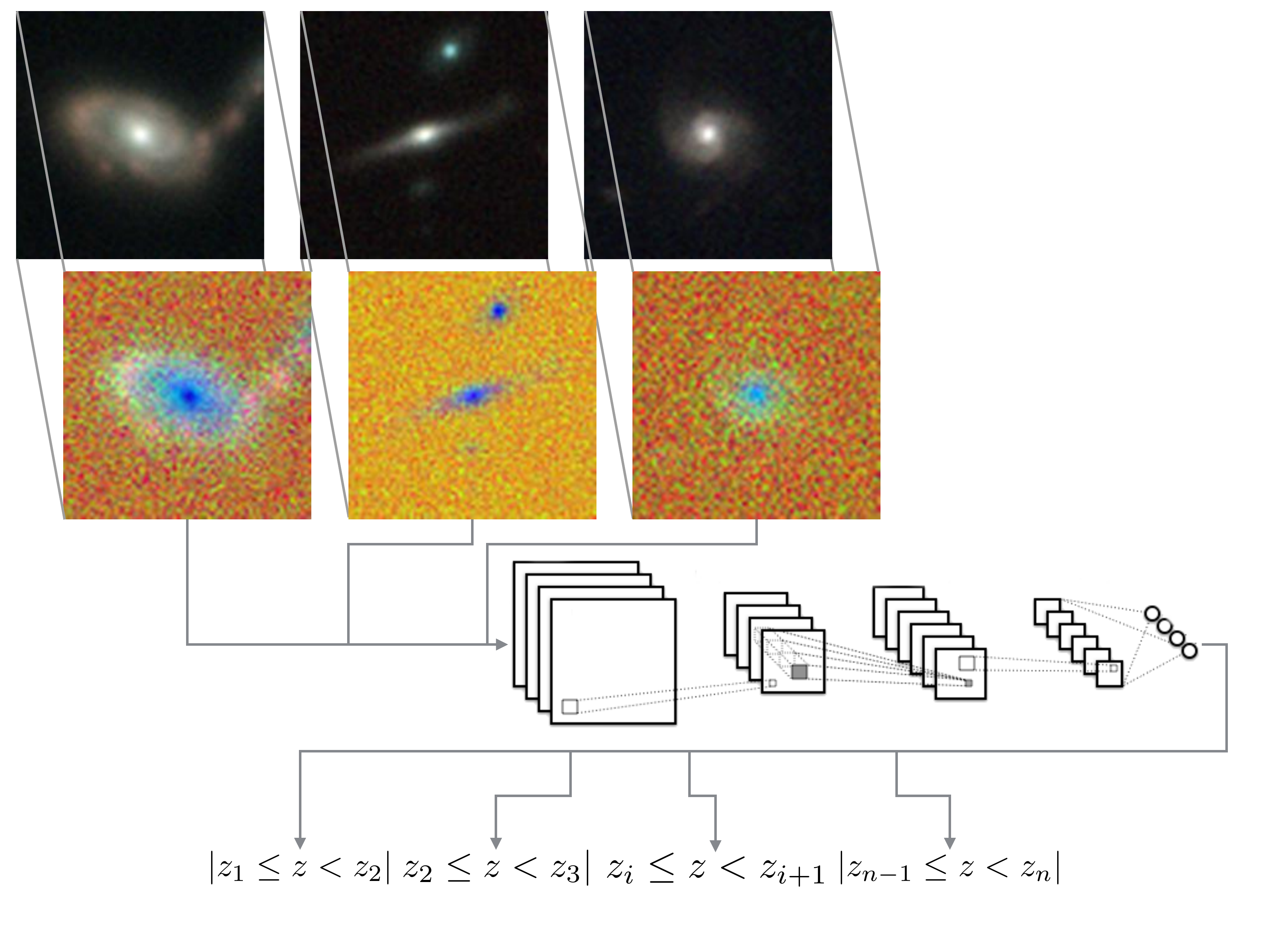}
\caption{The experimental set up with the ImageNet inspired Deep Neural Network (DNN) with Convolutional layers. We convert the pixel fluxes  (top images) to pixel magnitudes and subtract magnitudes to make pixel colours.  {\rr The following colours are placed into separate image layers, the $i-z$ color maps to the R layer pixels, $r-i$ to the G layer pixels, and $g-r$ to the B layer pixels. Finally we pass the $r$ band pixel magnitude into an additional Alpha layer to produce the RGBA image, as seen in the second row}. These images are passed into a DNN (illustrated by the third row) to predict the galaxy redshift ($z$) bin (bottom panel). Partial image credit in text.}
\label{DNN}
\end{figure}

In the top row of Fig. \ref{DNN} we show RGB jpeg images of three example galaxies with the following mappings{\rr; $g$ band magnitude $\rightarrow$ R, $r$ band  magnitude $\rightarrow$ G,  and  the $i$ band  magnitude $\rightarrow$ B. } {\rr All pixel  magnitudes are further} rescaled across the entire layer to be integers within the range 0 to 255 for viewing purposes only.  We further modify these base images to be more suitable for photometric redshift analysis by producing pixel colours from the pixel magnitudes and map pixel colours to each RGB layer pixel. We map the pixel colours $i-z$ to the R layer pixels, $r-i$ to the G layer pixels, and $g-r$ to the B layer pixels. Finally we pass the $r$ band pixel magnitude into an additional Alpha layer to produce an RGBA image. The $r$ band magnitude is often used in this way to act as a {\rr pivot point which provides an overall normalisation to the input data. This may be useful during training and is common practice in photometric redshift analysis using neural networks \cite[see e.g.,][]{2014A&A...568A.126B}}. Examples of these modified images are shown in the second row of Fig. \ref{DNN}, but we show only the RGB values for viewing purposes.

During the analysis we scale all of the images, such that the maximum pixel value of 255 corresponds to the largest value across all training and test images  in each of the RGBA layers separately. Likewise the minimum pixel value of 0 is set to be the smallest value in each layers across all images.


For a comparison with standard machine learning architectures we obtain model magnitudes measured by the SDSS photometric pipeline for each of the galaxies. To produce a fair comparison with the image analysis, we choose to use the de-reddened model magnitudes in the $g,r,i,z$ bands and the size of each galaxy measured by the Petrosian radius in the $r$ band.

\begin{figure}[ht]
\centering
\includegraphics[scale=0.6,clip=true,trim=0 10 40 30]{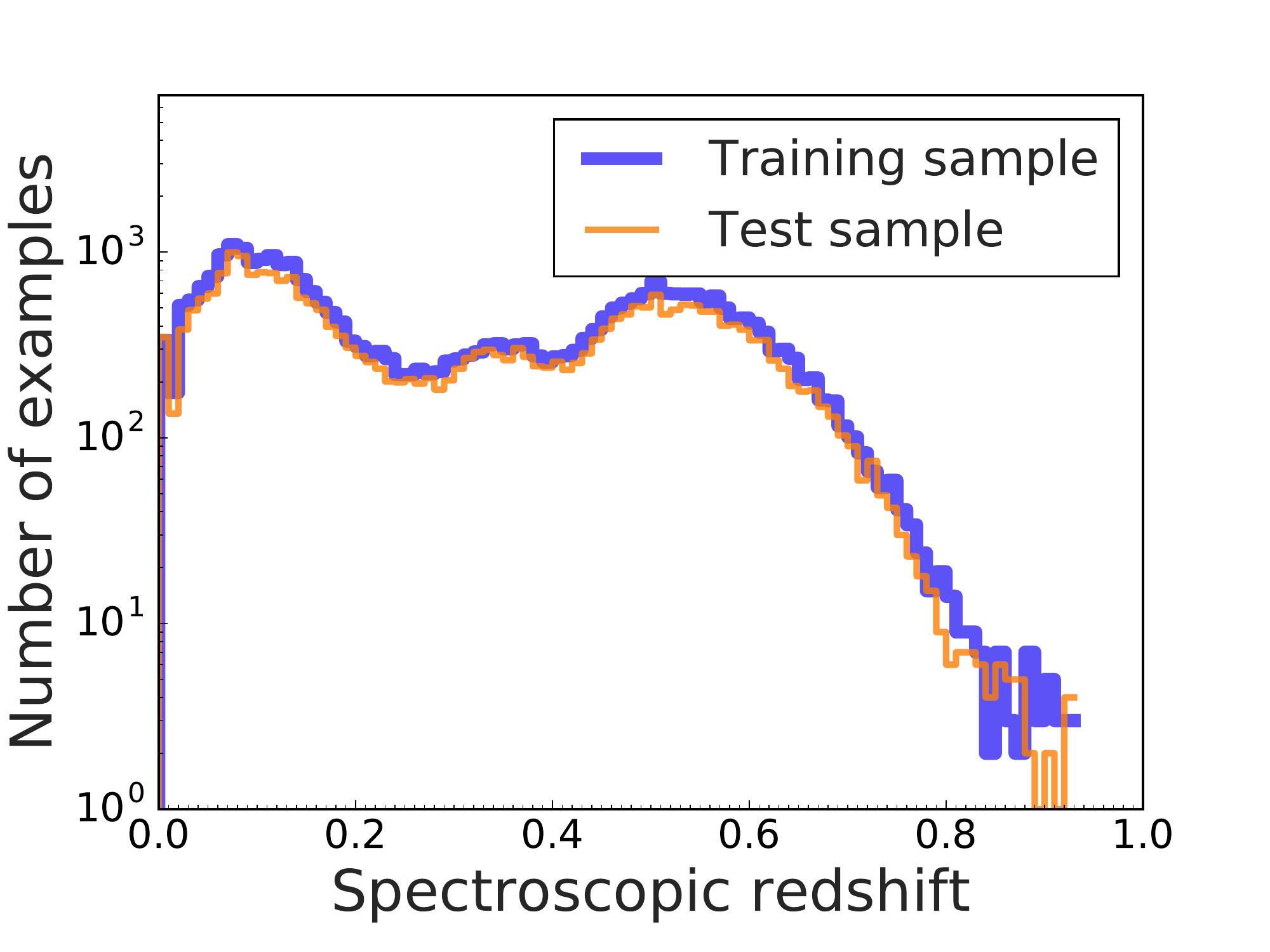}
\caption{The redshift number distribution of training  (thick blue line) and test (thin orange line) galaxies used in this work. The stepped lines represent the classification bins which are of width 0.01.}
\label{bins}
\end{figure}

We randomly shuffle and subdivide the {\rr 64,647} galaxies into training, cross-validation and test samples of size 33,167, 4,047, and {\rr 27,433}. In what follows we train the machine learning architectures on the training sample. We then vary the hyper-parameters of the machine learning architecture and retrain a new model. We select which is the best trained model using the cross-validation sample, {\rr which is completely independent from the training sample.} After choosing a final model, we pass the test sample through the final model to obtain machine learning redshift predictions. These redshift distributions produce {\rr a fair} estimate of the ability of the machine learning architecture to predict redshifts for other galaxies which are representative of the training sample.  In Fig. \ref{bins} we show the spectroscopic redshift number distribution of training  (thick blue line) and test (thin orange line) galaxies used in this work in. The stepped lines represent the classification bins which have a width of 0.01 in redshift. 

\section{Machine Learning Architectures}
\label{MLA}
In this work we utilise the latest developments in the field of machine learning by using Deep Neural Networks (DNNs). In particular we pass the entire galaxy image into the DNN to obtain a redshift estimate.  As a comparison method we use a machine learning framework called boosted trees which produce the current state of the art photometric redshift estimates using standard photometric features. We describe both architectures in more detail below.

\subsection{Deep Neural Networks}
Major advances in many areas of machine learning have recently been produced using DNNs. DNNs are based on standard neural networks, which are themselves inspired by the learning connections between biological neurons and synapses in the human brain. Neural networks have input layers, hidden layers and output layers. For our purposes, the input layers are the real valued photometric feature vectors that are measured for each galaxy. The output layer is the real valued floating point prediction for the redshift. The hidden layers are connected to the input layers and they combine and weight the input values to produce a new real valued number, which is then passed to the output layer. The weights of the connections between the layers are updated during the training phase in order to make the output value for each galaxy as close as possible to the spectroscopic redshift for that galaxy.

DNNs depart from these simple neural networks by constructing many hidden layers, with many multiple connected neurons per layer. DNNs can also accept images as input layers using an architecture called Convolutional Neural Networks\cite{lecun95convolutional}, instead of vectors of real valued numbers. The Convolutional Neural Networks retain information about the physical location of pixels with respect to other pixels and are used efficiently in combination with the \textsc{Max Out} algorithm\cite{2013arXiv1302.4389G}. The power of DNNs comes from recent advances in how the connections between the many millions of neurons are trained. Previously the many millions of connections would quickly overfit even large training sets, and thereby {\rr lose} the DNNs predictive power. One major advancement is the \textsc{Dropout}\cite{2012arXiv1207.0580H} technique, which ignores a random number of neurons during each training round. This effectively results in each training round learning a `weak model', which are then combined to produce a final model with a lot of predictive power, and a lower chance of overfitting. Weak models have low predictive power by themselves, however the predictions of many weak models can be weighted and combined to produce models with much stronger predictive power.

To further ensure that the DNN does not overfit we apply data augmentation techniques to produce many training examples for each of the original input images. We apply random image flipping and rotations, and randomly select a sub patch of size 60x60 pixels to pass into the DNN. {\rr The image rotations are performed in discreet 90 degree intervals. We use these methods to increase the training sample size by a factor of 80.} We do not currently apply whitening techniques to add noise to the images, which can further help with overfitting.

We choose to use a base DNN architecture inspired by \cite{NIPS2012_4824}  that obtains state of the art results on the ImageNet dataset\cite{2014arXiv1409.0575R}. We modify the base DNN to accept images of dimension 4x60x60 and which produces an output layer with 94 classification bins, which correspond to redshift slices of width 0.01. We have also explored a limited range of DNN architectures. For example we find {\rr that} using galaxy images of dimensions 4x32x32 reduces the performance by more than 30\%, and increasing the dropout fraction from 0.4 to 0.9 we find that a dropout fraction of 0.6 produces slightly higher accuracy on the cross-validation. In future work we will provide a more detailed analysis of the effect of varying the hyper-parameter choices for the DNN architecture. We describe the full DNN architecture in more detail in the appendix but note here than it contains some 23 layers. In this work we use the package \textsc{GraphLab}\cite{2010arXiv1006.4990L} as the main tool for building and training DNNs.

We show an illustration of the ImageNet inspired DNN with Convolutional Neural Network layers in the third row of Fig. \ref{DNN} which is an altered version of an image found on \url{http://deeplearning.net/tutorial/lenet.html}. The modified galaxy images  (second row panels) are passed into the ImageNet DNN (third row) to predict the galaxy redshift bin (final row) in a classification analysis. {\rr In Fig. \ref{bins} we present the distributions of the training and test data per each redshift classification bin.}

\subsection{Tree methods}
Once a galaxy has been observed and its photometric properties measured, it can be placed along with other galaxies into a high dimensional scatter diagram in which each dimension {\rr corresponds} to a chosen input feature.  Decision trees are machine learning architectures which subdivide this high dimensional space into high dimensional boxes. Each new split, or box, is chosen during the training phase to maximise the similarity of the spectroscopic redshifts for all galaxies which fall within the same box. Once the space has been suitably subdivided the training ends and each box is assigned a redshift estimate which is the mean value of all remaining galaxies within the box. Test data is then placed into the high dimensional space, and the machine learning redshift estimate is assigned to the test data from the value of the hyper-box which contains it.

One may think of each individual decision tree, or configuration of hyper-boxes, as learning a weak model, and the power of tree based methods {\rr comes} from combining the results of many weak models to produce a final model with strong predictive power and a low chance of over fitting. There exist many techniques to choose how the individual trees should be grown, and how the trees should be combined, one of which is called Adaptive boosting, or \textsc{AdaBoost}\cite{Freund1997119,Drucker:1997:IRU:645526.657132}. \textsc{AdaBoost} has recently been shown to provide the most accurate galaxy redshift estimates when compared with many other machine learning technologies \cite{2015arXiv150308214H}. The power of \textsc{AdaBoost} is due to the algorithm preferentially attempting to learn a good model, for those training examples with the worst performance in the previous training round. {\rr We note that other boosting algorithms exist, such as LogitBoost \citep[][]{friedman2000}, but have not been widely adopted by the astrophysics community \citep[however, see][]{zhangmining}.}

The hyper-parameters of the scikit-learn \cite{scikit-learn} implementation of \textsc{AdaBoost} with regression trees are the number of trees combined to make the final model, the minimum number of training examples  in the final  hyper-boxes, the loss function, and the learning rate. We explore the full range of loss functions and other hyper-parameters within the scikit-learn implementation of \textsc{AdaBoost}. For more details on combining trees with \textsc{AdaBoost} and for further descriptions of the hyper-parameters, we refer the reader to \cite{hastie01statisticallearning}. In what follows we refer to this standard machine learning architecture using the magnitudes, colours and a $r$ band Petrosian radius as `AdaBoost'

\section{Results}
\label{res}
We train both of the machine learning architectures (hereafter MLA) on the same sample of training galaxies, and determine how well each MLA has been trained by passing the cross-validation sample through the learnt machine. For DNNs we use the full galaxy image as an input, and for AdaBoost we use the measured magnitudes, colours and radii. The output of AdaBoost is the real valued number $z_{ML}$, that corresponds to the photometric redshift. The output of the DNN is the redshift bin that the classified galaxy is most likely to have. The DNN randomly extracts a sub image of size 4x60x60 from the original image of size 4x72x72 and therefore can produce a different redshift prediction for each random sampling of the same image. We therefore pass each galaxy image into the final DNN one hundred times to produce a redshift classification distribution, which we then convert to a redshift vector. We calculate the mean and standard deviation of this redshift vector and label the mean redshift for this galaxy  as  $z_{ML}$. We note that if we choose to use the median instead of the mean as the redshift estimate, the final statistics vary very little.

We construct the residual vector $\Delta_z = z_{ML}-z_{spec}$  which is the difference between the machine learning redshift and the spectroscopic redshift. We  measure the following metrics: $\mu,\sigma_{68},\sigma_{95}$, corresponding to the median value of $\Delta_{z}$, and the values corresponding to the 68\% and 95\% spread of $\Delta_{z}$. We additionally measure the `outlier rate' defined as fraction of galaxies for which $|\Delta_{z}/(1+z_{spec})|>0.15$.  If the residual distribution were described well by a Gaussian distribution, the choice of $\sigma_{68}$ would correspond to the standard deviation, and $\mu$ would be equivalent to the mean. However most photometric redshift residual distributions have longer tails and are more peaked than a Gaussian distribution and therefore the standard deviation is not representative of the dispersion of the data.

For AdaBoost we randomly explore the hyper-parameter space 500 times and select the trained machine with the lowest value of $\sigma_{68}$ as measured on the cross-validation set. Similarly, we select the final DNN from the handful of models that we explored,  to be the model with the lowest value of $\sigma_{68}$ as measured on the cross-validation set.

After deciding upon a final model for both MLAs we pass the sample of test galaxies, which is not used during training or model selection phase, through each MLA to obtain a final set of machine learning photometric redshifts. This represents an unbiased estimate of the ability of the MLAs to produce redshift estimates for other galaxies, however these galaxies must be similar to, or representative of, the training sample. We again construct the residual redshift vector and measure the same statistics as before.

\begin{figure}
\centering
\includegraphics[scale=0.6,clip=true,trim=10 10 47 30]{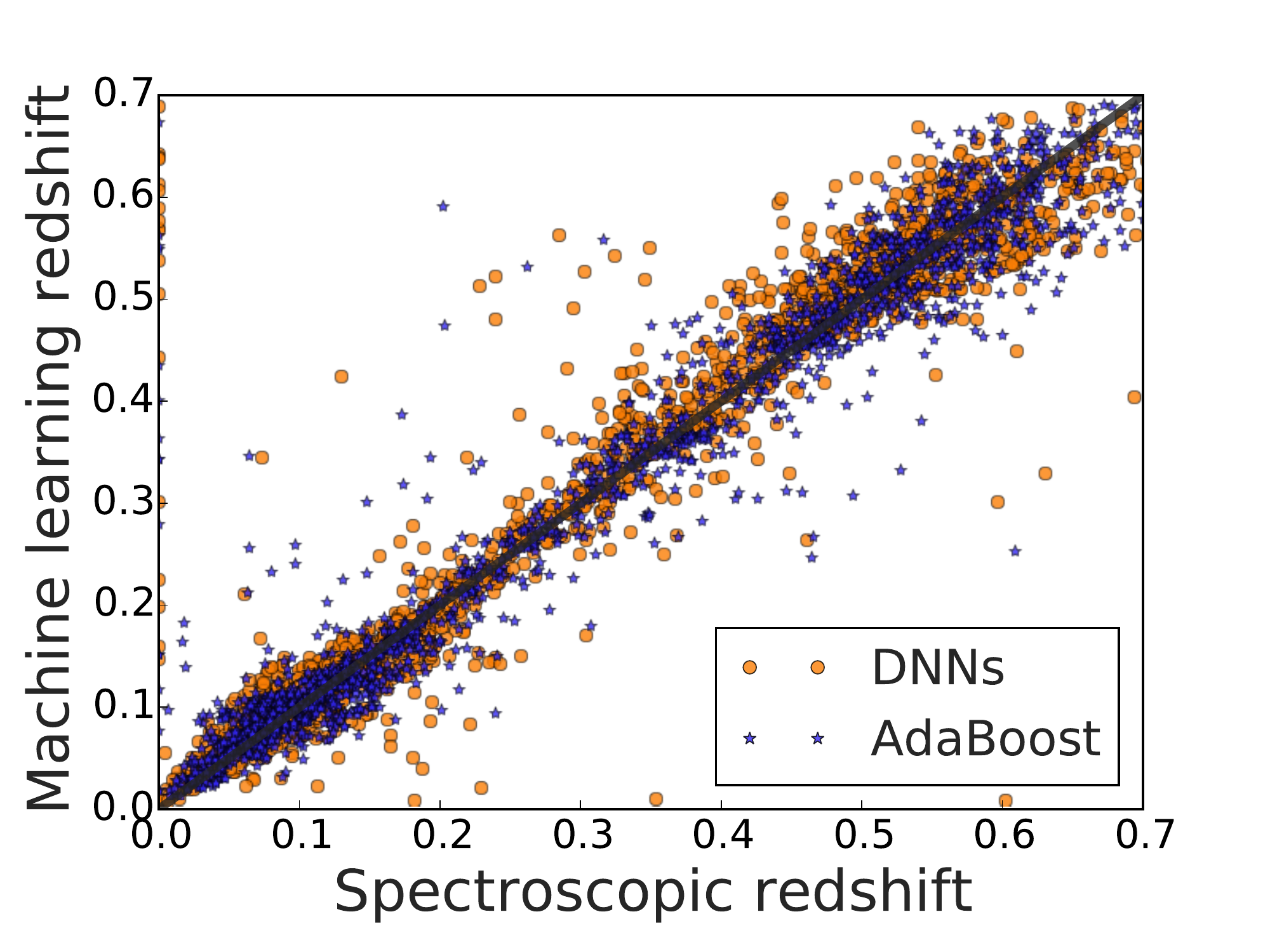}
\includegraphics[scale=0.6,clip=true,trim=10 10 47 30]{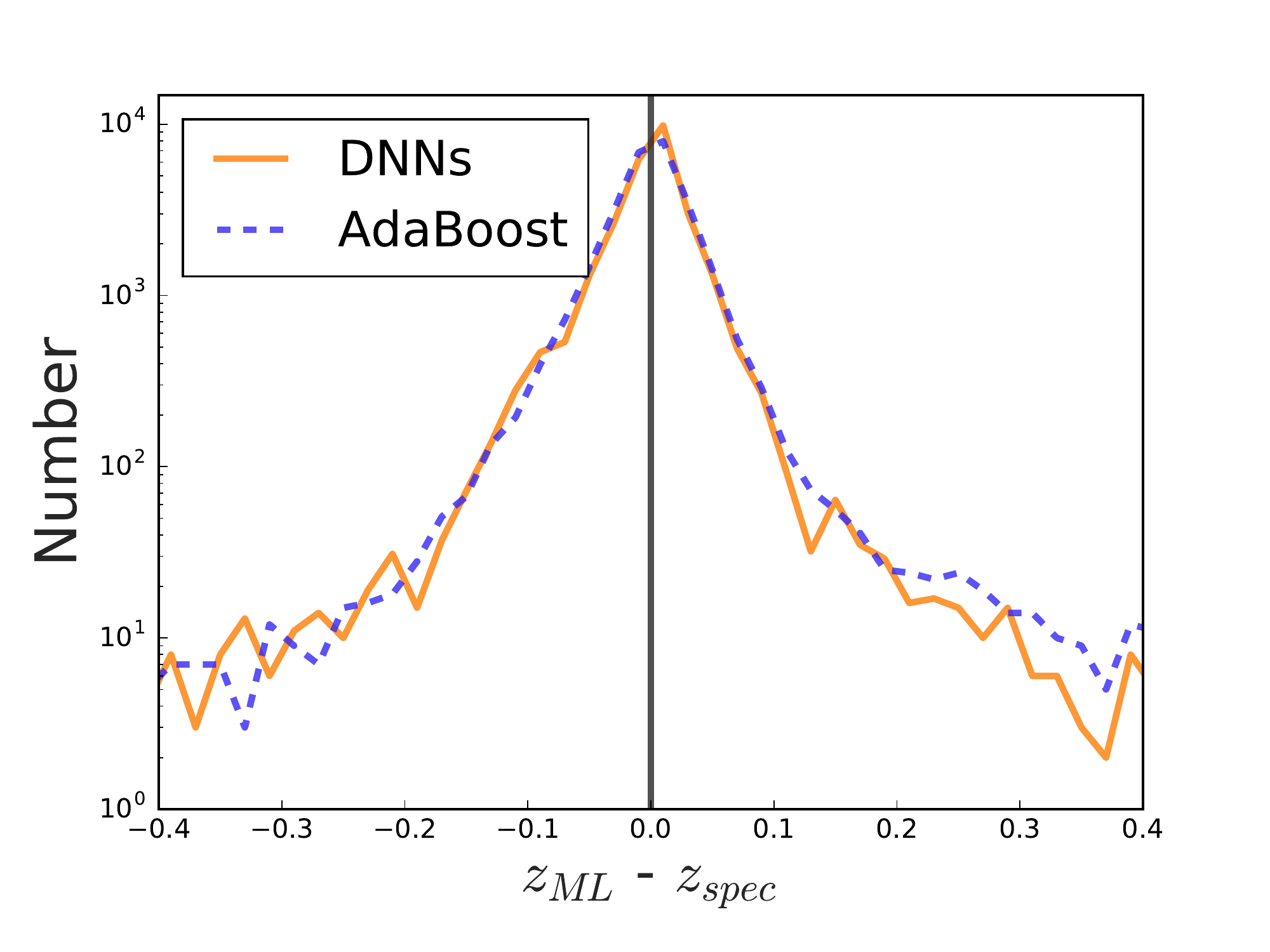}
\caption{The top panel shows the DNNs machine learning redshift  estimate against the spectroscopic redshift by the orange circles, and the AdaBoost machine learning redshift estimate by the blue stars. The bottom panel presents histograms of the redshift residuals for DNNs by the solid orange line, and AdaBoost by the blue dotted line. The dark grey solid line shows the line of equality in the top panel, and the line described by  $\Delta_z=0$ in the bottom panel.}
\label{DNNres}
\end{figure}

We present the results of the MLAs in Fig. \ref{DNNres}. The top panel shows a scatter plot of the DNN and AdaBoost redshift estimates against the spectroscopic redshift for each galaxy. The bottom panel shows histograms of the redshift residuals. We present the results using the DNNs by the orange circles and solid lines, and the AdaBoost results  by the blue stars symbols and dotted lines. The dark grey solid line shows the line of equality in the top panel, and the line described by $\Delta_z=0$ in the bottom panel.
We show the values of each of the measured statistics in Table \ref{photoRapter}. {\rr We highlight that the values of $\mu$ and $ \sigma_{68}$ for the DNNs are identical (to the quoted precision)} to those values obtained from AdaBoost. We find that the outlier fraction is larger by 10\% for the DNNs {\rr (1.71\%)} compared with AdaBoost (1.52\%).

\section{Discussion and conclusions}
\label{concl}
{\rr 
Robust photometric redshift estimates are a critical component of maximising the cosmological information content available from current and future photometric galaxy surveys. Indeed, recent work \cite{2015arXiv150308215R} show how the mis-estimation of the galaxy redshift distribution for a sample of galaxies produces biases in many correlation function analyses, and other work shows how these biases effect cosmology \citep[e.g.,][]{2015arXiv150705909B}.

Until now photometric redshifts have been estimated by first extracting quantities from the galaxy image which are deemed salient by the user. The extracted quantities are normally fluxes within a chosen aperture, or radii describing some aspect of the galaxy profile. The extracted quantities are then either compared to theoretical models of galaxy evolution, for example when using template based methods, or are used to learn the mapping between the measured quantities and the spectroscopic redshift for the subset of the data which already has redshifts, for example when using standard machine learning methods.

In this work we propose a completely new method to estimate photometric redshifts by passing the full galaxy imaging into a Deep Neural Network {\rr (DNNs)}. The main advantage of this method is that the user does not prejudice the choice of measured properties extracted from the galaxy image apriori.  

One can view this new approach as the most extreme form of {\it feature importance} possible \citep{2015MNRAS.449.1275H}. Feature importance ranks the chosen properties (or features) of the galaxy by their predictive power for the task at hand. In this approach features are not chosen a priori, but learnt during training. One consequence of this additional freedom is the massive increase in computational cost involved with this type of analysis, compared with a standard analysis using predefined features. It is therefore necessary to train the DNNs using codes optimised for GPUs, and such codes are becoming more widespread and user friendly, see e.g. GrapLab\footnote{dato.com}, Keras\footnote{keras.io}, or pylearn2\footnote{deeplearning.net/software/pylearn2}.

We compare our results using DNNs with a standard machine learning photometric redshift analysis using the machine learning algorithm called AdaBoost\cite{Freund1997119,Drucker:1997:IRU:645526.657132} and the following input features; the deredened model magnitudes $g,r,i,z$, colors derived from the magnitudes, and the $r$ band Petrosian radius. This standard machine learning architecture has recently been shown to produce state of the art photometric redshift estimates\cite{2015arXiv150308214H}. These choices of input features are made for maximal comparison with other current and future photometric surveys, for example the Dark Energy Survey \citep[][]{2005astro.ph.10346T}.

For the DNN analysis we obtain $r,g,i,z$ FITS images which we pre-process to generate four layer RGBA images, with the following mapping between layers and pixel colours and pixel magnitudes; the colours $i-z \rightarrow$  R layer, $r-i\rightarrow$ G layer and $g-r\rightarrow$ B layer. Finally we map the $r$ band pixel magnitude into Alpha layer of the RGBA image to provide a pivot point. The layers are further scaled to have integer values between 0 and 255, over the entire data sample.

One future extension of this work is to explore more realistic effects when using images with both artefacts and masked pixels, potentially due to survey boundaries, cosmic rays, or poor observing conditions. We find that none of the SDSS images used in this analysis have these problems. When using DNNs it is important to perform image rescaling, such that range of values do not span orders of magnitudes. Artefacts and masked pixels will therefore have to be dealt with carefully when they do occur.

We download the above photometric features and images for 64,647 galaxies from the SDSS website. We divide this data into a training, cross-validation and test sample of size 33,167, 4,047, and 27,433. We choose to build sample sizes which are relatively small compared to the full SDSS spectral data set because of the computational cost of obtaining images, training the DNN and obtaining predictions. Both the training and the prediction phases of the DNN experiment require approximately 5 orders or magnitude more computing resources than the standard analysis. This is a severe limitation of using the DNN method, especially because the obtained predictions are comparable to those obtained by the faster standard machine learning algorithms. However deep machine learning has made radical improvements and produces state of the art predictions when applied to a variety of tasks. We therefore expect that as computing resources increase, and a more exhaustive search of hyper-parameter settings is performed, the predictive power of DNNs may well improve over standard machine learning algorithms. Such alterations of the DNN architecture involve varying the number and shape of the convolutional neural network layers, the drop out fraction between the different layers, the number and size of the flattened hidden layers and their activation functions, and the output layers from a binned classification analysis to a regression analysis. One may further extract the outputs of the final hidden layer and use these as input features in a standard machine learning analysis. 

In this work we explore a limited number of different DNN architectures to select a good fitting model. We leave a full analysis of DNN architectures to future work and refer to the appendix for a fuller description of the DNN architecture used in this work.

We construct the residual vector $\Delta_z = z_{ML}-z_{spec}$  which is the difference between the machine learning photometric redshift $z_{ML}$ and the spectroscopic redshift. We  measure the following metrics: $\mu,\sigma_{68},\sigma_{95}$, corresponding to the median value of $\Delta_{z}$, and the values corresponding to the 68\% and 95\% spread of $\Delta_{z}$, and we additionally measure the `outlier rate' defined as fraction of galaxies for which $|\Delta_{z}/(1+z_{spec})|>0.15$. 

Other possible extensions to this work include the estimation of full galaxy redshift probability distribution functions (pdfs) instead of redshift point predictions. A starting point for this work is to follow that of \cite{2013arXiv1312.1287B}, who estimate redshift distributions for galaxies using neural networks. We expect that the estimation of pdfs will further marginally increase the computation cost of the analysis.

We note that the value of $\mu$ and $ \sigma_{68}$ for the DNNs (0.0, 0.03) are almost identical to those values obtained from AdaBoost (0.001, 0.03). We find that the outlier fraction is {\rr slightly} larger by {\rr 10\%} for the DNNs (1.71\%) compared with AdaBoost (1.56\%).

In future work we will extend this analysis to include more training and test galaxies from the SDSS and other datasets. We will also begin to explore a much larger range of DNN architectures, and other input image configurations.

}
\section*{Appendix: Deep Neural Network Archtitecture}
In what follows we describe the DNN used in this work. We note that this DNN is inspired by\cite{NIPS2012_4824} and further modified to suit both the input image shape choices and the output redshift classifications binning.

First the images of size 72x72x4 are pre-processed to obtain pixel colours,  which are mapped to the RGBA layers as described in the data section. We then extract random contiguous images of {\rr shape} 60x60x4  from the pre-processed images. These random images are passed into the first layer of the net which is a Convolution Layer (denoted by $C_{3,10}$) which itself applies a learning smoothing filter of size 3x3x4 into a new pixel value which is stored in new sub images in the the next layer. Ten such sub images are generated in this way. The next layer is a Rectified Linear Layer ($R$) which transforms all of the input values into output values using the function $f(x)=max(0,x)$. These values are then transformed by a MaxPooling Layer ($MP_{3}$) which is similar to the filtering in the $C$ layer, but instead outputs the maximum value of the 3x3 filtered sub image into the next layer. The next layer is a Local Renormalisation Layer ($RN_5$) which normalised the output values by the values coming from 5 neighbouring neurons. The subsequent Layers are $C_{5,256}\rightarrow$$R\rightarrow$ $MP_{3}\rightarrow$$RN_5\rightarrow$$C_{3,384}\rightarrow$$R\rightarrow$$C_{3,384}\rightarrow$$R\rightarrow$$C_{3,256}\rightarrow$$R,MP_{3}$, which is then followed by a flattening layer which converts the Convolutional type layers into flat layers such as those found in standard neural networks. {\rr The flattened layer is then followed by a Fully Connected ($F_{4096}$) layer with 4096  neurons followed by R and then a Dropout Layer $D_{0.6}$. The dropput layer transforms the incoming values by probabilistically ignoring them during training, with a probability of 0.6.} This Dropout layer is followed by $F_{4096}$, $R$, $F_{94}$ corresponding to the 94 redshift classes which are finally normalised and converted into class probabilities using a Softmax layer.

\section*{Appendix: MySQL data query}
{\rr
We select data from the SDSS CasJobs website by running the following MySQL query in the Data Release 10 context:
\begin{verbatim}
select p.objid, s.specobjid, s.ra, s.dec,
s.z as spec_z, s.zerr as err_spec_z,
p.dered_u,p.dered_g,p.dered_r,p.dered_i,p.dered_z,p.PETRORAD_R,
p.extinction_g, p.extinction_r,p.extinction_i,p.extinction_z
into mydb.DR10_DNN
from Specobjall s join photoPrimary p on (s.bestobjid =p.objid)  
and  p.deVRad_r >0 and p.deVRad_r<30 and 
p.dered_r>0 and p.dered_r < 22 and s.z>0 and s.z<2 and 
s.zerr>0 and s.zerr<0.1 and 
p.expRad_r>0 and p.expRad_r <30 and p.type=3
\end{verbatim}
This results in 1,918,221 galaxies, of which we randomly select 64,647 for use in this paper.
}
\section*{Acknowledgements}
I would like to thank Sander Deileman and Kerstin Peach for useful discussions and Jochen Weller and Stella Seitz for proof reading and comments, {\rr and an anonymous referee who has provided comments and feedback which have improved the quality and readability of the paper.} The author declares no competing financial interests.

\begin{table}
\centering
  \begin{tabular}{ | c | c | c | c | c | } \hline
 MLA  & $\mu $ & $ \sigma_{68}$ &  $ \sigma_{95}$ &  $|\Delta_{z}/(1+z_{spec})|>0.15$\\ \hline
DNNs &$0.00$ & $0.030$  & $0.10$ & $1.71$\% \\
AdaBoost & $-0.001$ & $0.030$  & $0.10$ & $1.56$\%  \\ \hline
  \end{tabular}
\caption{\label{photoRapter}  The statistics measured on each of the best machine learning architectures (MLA) are shown in the column headings, and are measured on the redshift residual distribution $\Delta_{z}$ of the test galaxies, which are not used during training or model selection. 
 }
\end{table}

\section*{References}
\bibliography{photoz}
\end{document}